\documentclass[pra,reprint,amsmath,amssymb,amsfonts,twocolumn,nofootinbib,showpacs]{revtex4}

\usepackage{bm,graphicx,mathrsfs}
\usepackage{graphicx}
\usepackage{epsfig}
\usepackage{amsmath,bbm}
\usepackage{amsfonts,amssymb}
\usepackage{times}
\usepackage{dsfont}

\newcommand{\be}{\begin{equation*}}
\newcommand{\ee}{\end{equation*}}
\newcommand{\bea}{\begin{eqnarray}}
\newcommand{\eea}{\end{eqnarray}}

\newcommand{\ket}[1]{|#1\rangle}
\newcommand{\bra}[1]{\langle#1|}

\def\>{\rangle}
\def\<{\langle}

\begin{document}

\title{Effective operator formalism for open quantum systems}
\author{Florentin Reiter\footnote{reiter@nbi.dk} and Anders S. S\o rensen\footnote{anders.sorensen@nbi.dk}}
\affiliation{QUANTOP, Danish Quantum Optics Center, Niels Bohr Institute, Blegdamsvej 17, DK-2100 Copenhagen \O, Denmark}
\date{\today}

\begin{abstract}
We present an effective operator formalism for open quantum systems. Employing perturbation theory and adiabatic elimination of excited states for a weakly driven system, we derive an effective master equation which reduces the evolution to the ground-state dynamics. The effective evolution involves a single effective Hamiltonian and one effective Lindblad operator for each naturally occurring decay process. Simple expressions are derived for the effective operators which can be directly applied to reach effective equations of motion for the ground states. We compare our method with the hitherto existing concepts for effective interactions and present physical examples for the application of our formalism, including dissipative state preparation by engineered decay processes.
\end{abstract}
\pacs{03.65.Ca, 03.65.Yz, 42.50.Lc, 03.65.Fd}
\maketitle

\section{Introduction}

Understanding the dynamics of open quantum systems \cite{Zoller} requires methods to model their temporal dynamics. Generally the evolution of an open quantum system is described by the master equation which determines the evolution of the system's density matrix $\rho$. Solving the master equation for the full density matrix is in many situations cumbersome. This is also the case for closed systems where the evolution can be described by wave functions. Since in general an open system involves both unitary and dissipative dynamics, the complexity of its evolution is, however, substantially higher than for the corresponding closed system, as the number of entries in the density matrix is the square of the number of entries in the wave function. To make the description of an open quantum system manageable as well as to gain physical insight into the evolution of the system it is therefore desirable to develop effective theories which reduce the complexity of the system. 

In an open system there are often quite different time scales associated with different effects, such that the Hilbert space can be divided into two parts, one for the rapidly decaying (excited) states, and one for the comparably stable (ground) states. For instance, for weakly driven atoms the evolution and decay of the excited states happen on a time scale which is fast compared to any other time scale in the system. In such situations it is desirable to eliminate the rapidly evolving excited states to get a simpler description of the slow evolution of the ground states. A standard method for doing this is adiabatic elimination \cite{Brion}, where the density matrix equations involving the excited states are solved by assuming a slow evolution of the ground states. This can then be used to describe an effective evolution of the ground states. This procedure can, however, be rather involved, as there are many density matrix elements if the system is large.

In this paper we present a simple method to eliminate the excited states and to reduce the system dynamics to the ground states. The method we present is essentially equivalent to adiabatic elimination but is much easier to apply in practice. By formalizing the procedures leading to adiabatic elimination we obtain simple expressions for the effective operators describing the ground-state evolution. With our expressions one avoids the often tedious steps leading to adiabatic elimination and can obtain the effective operators by evaluating simple formulas. In particular, we have found that these methods are very convenient for studying dissipative state preparation \cite{VWC, Kraus, Diehl, KRS, RKS}, where the goal is to engineer decay processes such that a system evolves into a desired state. To this end it is highly desirable to have a convenient tool to rapidly identify the effective dissipative dynamics of the system.

For closed systems with purely unitary couplings similar simplifications have been used to derive effective Hamiltonians, in particular by James and co-workers \cite{James1, James2, SorensenMolmer}.
If decoherence is added to the system, the joint unitary and dissipative dynamics can be captured by introducing non-Hermitian descriptions. The use of complex energies allows for the combination of the energy of a resonance with its width \cite{Gamow}. Correspondingly, non-Hermitian Hamiltonians are commonly used to describe the dynamics of open systems \cite{Molmer, Carmichael, Persson, Volya, Alhassid}. In quantum optics the use of non-Hermitian Hamiltonians is put in a more rigorous form by the so-called quantum jump formalism, or Monte Carlo wave function method \cite{Molmer}, which is equivalent to the evolution by a Markovian master equation. In this method the non-Hermitian Hamiltonian describes the evolution in the absence of decay, whereas quantum jumps are introduced at random times to account for the resulting state after a decay.

The separation of the Hilbert space into rapidly and slowly evolving ground and excited states, similar to what we consider here, has been studied also for non-Hermitian Hamiltonians. 
In particular, for a coupling of the ground to the excited states much weaker than the evolution inside the subspaces, a formalism for effective processes is provided by the Feshbach projection-operator approach \cite{Feshbach}. Similar formalisms for successive nuclear reactions have been studied by Weidenm\"uller and co-workers \cite{Weidenmuller}. Methods based on the Feshbach projection operator method have been used in several fields \cite{Rotter}. However, these treatments are only concerned with the evolution from the effective non-Hermitian Hamiltonian and therefore ignore the quantum jumps describing the state after the decay. The procedure we present here is an extension of the Feshbach projection-operator approach to also include these quantum jumps. As a result, our formalism can be used to describe the full evolution of the density matrix of the system after elimination of the excited states. This generalization is crucial for describing situations where we are also interested in the state of the system after a decay.

\begin{figure}[t]
\centering
\includegraphics[width=8.6cm]{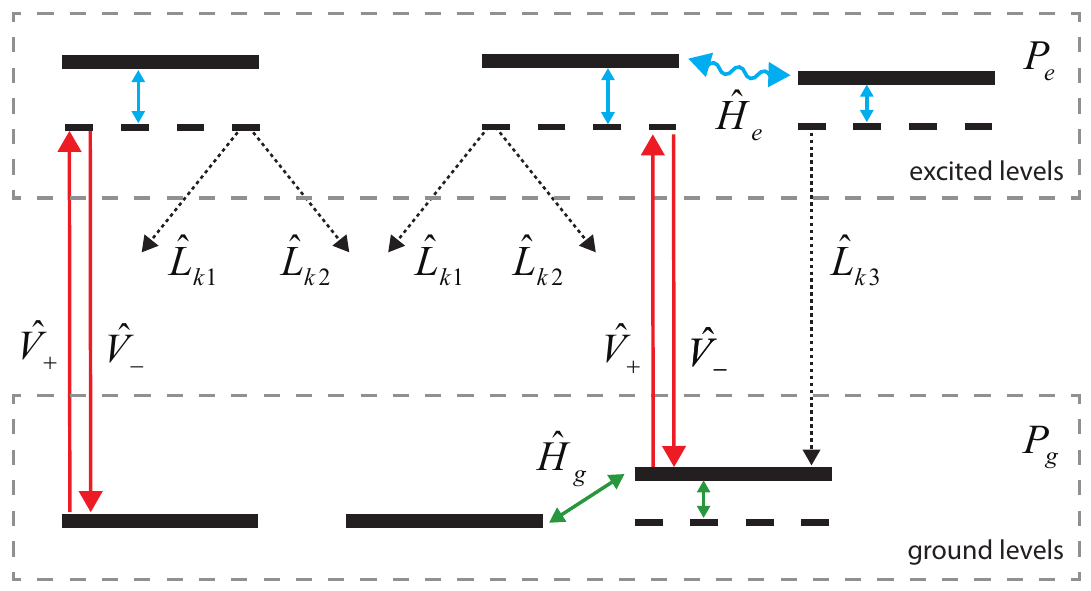}
\caption{(Color online) Ground and excited subspaces and couplings. The nondecaying ground states (corresponding to projector $P_{\rm g}$) are coupled to the decaying excited states (projector $P_{\rm e}$) by the perturbative (de-) excitations $\hat{V}_+$ ($\hat{V}_-$) [solid lines connecting the subspaces (red)]. The Lindblad operators $\hat{L}_k$ represent various decay process (dotted lines). The couplings inside the ground and excited subspaces are given by $\hat{H}_{\rm g}$ [solid lines inside the lower subspace (green)] and $\hat{H}_{\rm e}$ [solid lines inside the upper subspace (blue)], respectively.}
\label{FigSubspaces}
\end{figure}

\section{Aim of this paper}
Before proceeding we first outline the main result of this paper. As illustrated in Fig. \ref{FigSubspaces}, we assume the open system to consist of two distinct subspaces, one for the ground states and one for the decaying excited states. The couplings of these two subspaces are assumed to be perturbative. Furthermore, we assume that the dynamics of the system is Markovian such that the time evolution of the density operator $\rho$ can be described by a master equation of Lindblad form
\begin{align}
\label{master}
\dot{\rho} = -i \left[\hat{H}, \rho\right] + \sum_k \hat{L}_k \rho \hat{L}_k^{\dagger} - \frac{1}{2}\left(\hat{L}_k^{\dagger} \hat{L}_k \rho + \rho \hat{L}_k^{\dagger} \hat{L}_k\right), 
\end{align}
where $\hat{H}$ is the Hamiltonian of the system and each of the Lindblad operators $\hat{L}_k$ represents a source of decay which we assume to take the system from the excited to the ground subspace. 
By combining perturbation theory of the density operator and adiabatic elimination of the excited states we reduce the dynamics to an effective master equation involving only the ground-state manifold
\begin{align}
\label{effectivemaster}
\dot{\rho} = &-i \left[\hat{H}_{\rm eff}, \rho\right] + \sum_k \hat{L}_{\rm eff}^k \rho (\hat{L}_{\rm eff}^k)^{\dagger} - \nonumber \\ &- \frac{1}{2}\left((\hat{L}_{\rm eff}^k)^{\dagger} \hat{L}_{\rm eff}^k \rho + \rho (\hat{L}_{\rm eff}^k)^{\dagger} \hat{L}_{\rm eff}^k\right) 
\end{align}
with effective Hamilton and Lindblad operators
\begin{align}
\label{effectivehamilton2}
\hat{H}_{\rm eff} &= - \frac{1}{2} \hat{V}_- \left( \hat{H}_{\rm NH}^{-1} + (\hat{H}_{\rm NH}^{-1})^{\dagger} \right) \hat{V}_+ + \hat{H}_{\rm g}\\
\label{effectivelindblad2}
\hat{L}_{\rm eff}^k &= \hat{L}_k \hat{H}_{\rm NH}^{-1} \hat{V}_+.
\end{align}
connecting only the ground states. Here $\hat{V}_+$ ($\hat{V}_-$) are the perturbative (de-) excitations of the system and $\hat{H}_{\rm g}$ is the ground-state Hamiltonian. $\hat{H}_{\rm NH}$ is the non-Hermitian Hamiltonian of the quantum jump formalism
\begin{equation}
\label{hnh}
\hat{H}_{\rm NH} = \hat{H}_{\rm e} - \frac{i}{2} \sum_k \hat{L}_k^{\dagger} \hat{L}_k,
\end{equation}
with $\hat{H}_{\rm e}$ being the Hamiltonian in the excited-state manifold. The effective master equation of Eq. (\ref{effectivemaster}) provides an approximation of the dynamics in Eq. (\ref{master}) by the effective dynamics of its ground states. Thus, the effective operator formalism allows for a substantial reduction of the complexity of the dynamics of an open system. In essence, adiabatic elimination of the excited states in the presence of both coherent and dissipative processes is formalized in a compact form by the effective operators of Eqs. (\ref{effectivehamilton2}) and (\ref{effectivelindblad2}).

In the following section, Sec. \ref{SectionDerivation}, we will derive the effective operators of Eqs. (\ref{effectivehamilton2}) and (\ref{effectivelindblad2}). In Sec. \ref{Two} we show a first elementary application of our formalism to a driven dissipative two-level system. Thereupon, we discuss the possibility to use our effective operator formalism to engineer decay processes in a four-level system in Sec. \ref{SectionEngineered}.

In Sec. \ref{SectionGround} we turn to an extension of the formalism which includes nonperturbative ground-state couplings. This extended formalism will then be used for a detailed analysis of the effective processes in a three-level Raman system in Sec. \ref{SectionRaman}.
In Sec. \ref{SectionFields} we introduce a second extension of the effective operator formalism that allows for several perturbations or fields. The most general formalism is presented in Sec. \ref{SectionGeneral}. A comparison to similar existing methods is provided in Sec. \ref{SectionComparison}.
Readers who are more interested in the applications than in the derivations may turn to Secs. \ref{Two}, \ref{SectionEngineered}, and \ref{SectionRaman}, where we give examples of effective coherent and dissipative processes in typical quantum systems and discuss a simple dissipative state preparation scheme.

\section{Derivation of the effective operators}
\label{SectionDerivation}
We now present the derivation of the effective equation of motion in Eqs. (\ref{effectivemaster})--(\ref{effectivelindblad2}). The evolution of the density operator $\rho(t)$ in the Schr\"odinger picture is governed by the master equation of Lindblad form, given by Eq. (\ref{master}). The Hamiltonian $\hat{H}$ stands for unitary couplings of the system, such as coherent driving. Each Lindblad ``jump" operator $\hat{L}_k$ accounts for a dissipative process, such as spontaneous emission.

\subsection{Projection-operator formalism}
We use the projection-operator method of Feshbach \cite{Feshbach} to structure the Hilbert space into two subspaces, one for the ground states and one for the excited states, represented by the projection operators $P_{\rm g}$ and $P_{\rm e}$, with $P_{\rm g}+P_{\rm e}=\mathds{1}$ and $P_{\rm g} P_{\rm e}=0$. Accordingly, we divide the Hamiltonian into four parts:
\begin{align}
\label{Hamiltonian}
\hat{H} &= \hat{H}_{\rm g}+\hat{H}_{\rm e}+\hat{V}_++\hat{V}_-. 
\end{align}
Here, the interactions inside the ground subspace are labeled by $\hat{H}_{\rm g} \equiv P_{\rm g} \hat{H} P_{\rm g}$, and inside the excited subspace by $\hat{H}_{\rm e} \equiv P_{\rm e} \hat{H} P_{\rm e}$. The perturbative excitations $\hat{V}_+ \equiv P_{\rm e} \hat{H} P_{\rm g}$ and deexcitations $\hat{V}_- \equiv P_{\rm g} \hat{H} P_{\rm e}$ ($\hat{V}_+^\dagger = \hat{V}_-$ and $\hat{V} = \hat{V}_+ + \hat{V}_-$) connect the two subspaces.

We assume the ground states as stable and the excited states to be decaying to the ground states. The Lindblad operators can then always be written as $\hat{L}_k= P_{\rm g} \hat{L}_k P_{\rm e}$. The mentioned couplings inside and between the subspaces are illustrated in Fig. \ref{FigSubspaces}.

\subsection{Non-Hermitian time evolution in the quantum jump picture}
Combining unitary and dissipative dynamics within a single non-Hermitian Hamiltonian has widely been studied in various areas of physics, as mentioned in the Introduction. In quantum optics the use of non-Hermitian Hamiltonians is formalized by the so-called quantum jump picture \cite{Molmer}, in which an effective non-Hermitian Hamiltonian describes the evolution of the system in the absence of a quantum jump. In order to distinguish the non-Hermitian Hamiltonian of the excited states from the effective (Hermitian) Hamiltonian of Eq. (\ref{effectivehamilton2}) we have denoted it as $\hat{H}_{\rm NH}$ in Eq. (\ref{hnh}). It incorporates the excited state Hamiltonian $\hat{H}_{\rm e}$ and the decay terms of the anticommutator part of the master equation (\ref{master}). Introducing $\hat{H}_{\rm NH}$ to Eq. (\ref{master}) we obtain a reduced master equation
\begin{align}
\label{reducedlindblad}
\dot{\rho} = &- i \left((\hat{H}_{\rm NH} + \hat{H}_{\rm g} + \hat{V}) \rho - \rho (\hat{H}_{\rm NH}^{\dagger} + \hat{H}_{\rm g} + \hat{V})\right) \nonumber \\ &+ \sum_k \hat{L}_k \rho \hat{L}^{\dagger}_k
\end{align}
Here, we have included the decay terms which describe the loss of population from the excited states to $\hat{H}_{\rm NH}$ in the commutatorlike Hamiltonian part. The last ``feeding term" which describes the gain of the population of the ground states by decay from the excited states remains.

For ground-state interactions $\hat{H}_{\rm g}$ much weaker than those between the excited states $\hat{H}_{\rm e}$, the dynamics of the decaying excited states are mainly governed by the non-Hermitian Hamiltonian $\hat{H}_{\rm NH}$. As all excited states are decaying, all eigenvalues of $\hat{H}_{\rm NH}$ are nonzero so that its inverse $\hat{H}_{\rm NH}^{-1}$ exists within the excited-state subspace.

\subsection{Perturbation theory in the interaction picture}
In the following we assume the couplings of the ground and excited subspaces $\hat{V}_\pm$ to be sufficiently weak to be described as perturbations of the evolution governed by an unperturbed Hamiltonian $\hat{H}_0 \equiv \hat{H}_{\rm g} + \hat{H}_{\rm NH}$. Based on this assumption we perform perturbation theory of the density operator. To this end we change into the interaction picture by a transformation with the operator
\begin{equation}
\label{transform}
\hat{O}(t) = e^{-i \hat{H}_0 t} = e^{-i (\hat{H}_{\rm NH} + \hat{H}_{\rm g}) t}.
\end{equation}
Then the reduced master equation of Eq. (\ref{reducedlindblad}) transforms into
\begin{align}
\label{reducedlindblad2}
\dot{\tilde{\rho}}(t) = &-i\left(\tilde{V}(t) \tilde{\rho}(t) - \tilde{\rho}(t) \tilde{V}^{\dagger}(t)\right) + \sum_k \tilde{L}_k(t)\tilde{\rho}(t)\tilde{L}^{\dagger}_k(t)
\end{align}
with operators transformed accordingly
\begin{align}
\label{pertH}
\tilde{\rho}(t) &= \hat{O}^{-1}(t) \rho (\hat{O}^{-1}(t))^{\dagger}, \\
\tilde{V}(t) &= \hat{O}^{-1}(t) (\hat{H}_0 + \hat{V}) \hat{O}(t) + i \frac{d \hat{O}^{-1}}{dt} \hat{O}(t) \nonumber \\
&= \hat{O}^{-1}(t) \hat{V} \hat{O}(t), \\
\tilde{L}_k(t) &= \hat{O}^{-1}(t) \hat{L}_k \hat{O}(t).
\end{align}
To derive the effective operators we perform a perturbative expansion of the density operator in a small parameter $\epsilon$
\begin{align}
\tilde{\rho}(t) = \frac{1}{N}(\tilde{\rho}^{(0)}(t) + \epsilon \tilde{\rho}^{(1)}(t) + \epsilon^2 \tilde{\rho}^{(2)}(t) + ...)
\end{align}
and obtain a recursive formulation of the reduced master equation in powers of $\epsilon$,
\begin{align}
\label{recursion}
\dot{\tilde{\rho}}^{(n)}(t) = &-i (\tilde{V}(t) \tilde{\rho}^{(n-1)}(t) - \tilde{\rho}^{(n-1)}(t) \tilde{V}^{\dagger}(t)) + \\ &+ \sum_k \tilde{L}_k (t) \tilde{\rho}^{(n)}(t) \tilde{L}_k^{\dagger} (t),
\end{align}
where we have used that $\hat{V}$ is a small parameter $\hat{V} \propto \epsilon$. The first three orders of the recursive reduced master equation read
\begin{align}
\dot{\tilde{\rho}}^{(0)}(t) = &\sum_k \tilde{L}_k(t) \tilde{\rho}^{(0)}(t) \tilde{L}_k^{\dagger}(t), \\
\dot{\tilde{\rho}}^{(1)}(t) = &-i \left(\tilde{V}(t) \tilde{\rho}^{(0)}(t) - \tilde{\rho}^{(0)}(t)
 \tilde{V}^{\dagger}(t)\right) + \nonumber \\ &+ \sum_k \tilde{L}_k(t) \tilde{\rho}^{(1)}(t) \tilde{L}^{\dagger}_k(t), \\
\dot{\tilde{\rho}}^{(2)}(t) = &-i \left(\tilde{V}(t) \tilde{\rho}^{(1)}(t)- \tilde{\rho}^{(1)}(t) \tilde{V}^{\dagger}(t)\right) + \nonumber \\ &+ \sum_k \tilde{L}_k(t) \tilde{\rho}^{(2)}(t) \tilde{L}^{\dagger}_k(t).
\end{align}
In the absence of initial excitations, decay processes can be neglected for orders $n \leq 1$ so that
\begin{align}
\dot{\tilde{\rho}}^{(0)}(t) = &0, \\
\label{orderone}
 \dot{\tilde{\rho}}^{(1)}(t) = &-i \left(\tilde{V}(t) \tilde{\rho}^{(0)}(t) - \tilde{\rho}^{(0)}(t)
 \tilde{V}^{\dagger}(t)\right), \\
\label{ordertwo}
\dot{\tilde{\rho}}^{(2)}(t) = &-i \left(\tilde{V}(t) \tilde{\rho}^{(1)}(t)- \tilde{\rho}^{(1)}(t) \tilde{V}^{\dagger}(t)\right) + \nonumber \\ &+ \sum_k \tilde{L}_k(t) \tilde{\rho}^{(2)}(t) \tilde{L}^{\dagger}_k(t).
\end{align}
We use the projection operator approach for the density operator to separate the evolution of ground and excited states. In doing so we reduce the evolution of the ground states to
\begin{align}
P_{\rm g} \dot{\tilde{\rho}}^{(0)}(t) P_{\rm g} = &P_{\rm g} \dot{\tilde{\rho}}^{(1)}(t) P_{\rm g} = 0, \\
P_{\rm g} \dot{\tilde{\rho}}^{(2)}(t) P_{\rm g} = &-i P_{\rm g} \left(\tilde{V}(t) \tilde{\rho}^{(1)}(t)- \tilde{\rho}^{(1)}(t) \tilde{V}^{\dagger}(t)\right) P_{\rm g} + \nonumber\\ 
\label{masterground}
&+ \sum_k \tilde{L}_k(t) P_{\rm e} \tilde{\rho}^{(2)}(t) P_{\rm e} \tilde{L}^{\dagger}_k(t).
\end{align}
In the last line we have used that for each Lindblad operator we can write $\tilde{L}_k=P_{\rm g} \tilde{L}_k P_{\rm e}$, as decay only occurs from the excited to the ground states. Consequently, the ground states are connected by unitary and dissipative processes of second order. Also note that since the transformation in Eq. (\ref{transform}) is nonunitary, the perturbation $\tilde{V}(t)$ is non-Hermitian. For the dynamics of the excited states we find
\begin{align}
&P_{\rm e} \dot{\tilde{\rho}}^{(0)}(t) P_{\rm e} = P_{\rm e} \dot{\tilde{\rho}}^{(1)}(t) P_{\rm e} = 0\\
\label{masterexcited}
&P_{\rm e} \dot{\tilde{\rho}}^{(2)}(t) P_{\rm e} = - i P_{\rm e} \left( \tilde{V}(t) \tilde{\rho}^{(1)}(t) - \tilde{\rho}^{(1)}(t) \tilde{V}^\dagger(t) \right) P_{\rm e}.
\end{align}
As we have assumed that the excited states do not gain population from decay, Eq. (\ref{masterexcited}) does not exhibit any dissipative feeding terms. Hence, the evolution of the excited unitary dynamics is solely driven by the interaction Hamiltonian $\tilde{V}(t)$. While the dynamics of the second-order terms connect the states either in the ground or in the excited subspace we note that interactions between the subspaces are given by the first-order terms $P_{\rm g} \dot{\tilde{\rho}}^{(1)}(t) P_{\rm e}$ and $P_{\rm e} \dot{\tilde{\rho}}^{(1)}(t) P_{\rm g}$.

\subsection{Adiabatic elimination of the excited states}
In principle, a solution to the remaining second-order master equations for the ground and the excited states in Eqs. (\ref{masterground}) and (\ref{masterexcited}) can be computed. This solution can, however, still be very complicated. In particular, if the decaying excited states are almost unpopulated, it is preferable to obtain a more comprehensible solution. In the following, we choose to reduce the complexity of the dynamics by restricting it to the ground states. To this end, we perform adiabatic elimination of the excited states:
\begin{align}
P_{\rm e} \dot{\tilde{\rho}}^{(2)}(t) P_{\rm e} \approx 0.
\end{align}
Consequently, the dynamics of second order in Eq. (\ref{ordertwo}) are approximated by the dynamics of the ground states given by Eq. (\ref{masterground}). Below we follow the recursion of the perturbative expansion and carry out the perturbation integrals.

We obtain $P_{\rm e} \tilde{\rho}^{(2)}(t) P_{\rm e}$ by integrating Eq. (\ref{masterexcited}), and $\tilde{\rho}^{(1)}(t)$ by integrating Eq. (\ref{orderone}), and insert the resulting expressions into Eq. (\ref{masterground}). Having excluded the dynamics of the excited states by adiabatic elimination, we find the open system to evolve according to 
\begin{widetext}
\begin{align}
\label{pluggedin}
P_{\rm g} \dot{\tilde{\rho}}^{(2)}(t) P_{\rm g} =
&- P_{\rm g} \tilde{V}(t) \left(\int^t_0 dt' \ \tilde{V}(t') \tilde{\rho}^{(0)}(t') \right) P_{\rm g} - P_{\rm g} \left(\int^t_0 dt' \ \tilde{\rho}^{(0)}(t') \tilde{V}^\dagger(t') \right) \tilde{V}^\dagger(t) P_{\rm g} + \nonumber\\
&+ P_{\rm g} \sum_k \tilde{L}_k(t) P_{\rm e} \int_0^t dt' \int_0^{t'} dt'' \left(\tilde{V}(t') \tilde{\rho}^{(0)}(t'') \tilde{V}^{\dagger} (t'') + \tilde{V}(t'') \tilde{\rho}^{(0)}(t'') \tilde{V}^{\dagger}(t') \right) P_{\rm e} \tilde{L}^\dagger_k(t) P_{\rm g}.
\end{align}
\end{widetext}
Here, we have omitted terms where the density operator is sandwiched between perturbations $P_{\rm g} \tilde{V}$ and $\tilde{V} P_{\rm g}$. As $\tilde{\rho}^{(0)}$ lives in the ground-state subspace, these terms do not contribute to the ground-state evolution and can therefore be neglected. The remaining expression in Eq. (\ref{pluggedin}) contains two Hamiltonian-like and two Lindblad-like terms, for which we will carry out the integrals:
\begin{align}
I_1 &\equiv P_{\rm g} \tilde{V}(t) \int_0^t dt' \ \tilde{V}(t') \tilde{\rho}^{(0)}(t') P_{\rm g} \\
I_2 &\equiv P_{\rm e} \int_0^t dt' \int_0^{t'} dt'' \ \tilde{V}(t') \tilde{\rho}^{(0)} (t'') \tilde{V}^{\dagger}(t'') P_{\rm e}.
\label{integral2}
\end{align}
In this section we assume the direct interactions within the ground-state subspace to be perturbative. Hence, the ground-state evolution is negligibly small compared to the one for the excited states so that we have $\hat{O}(t) P_{\rm g} \simeq P_{\rm g}$. Consequently, $I_1$ simplifies to
\begin{align}
I_1 \approx \hat{V}_- \hat{O}(t) \left(\int_0^t dt' \ \hat{O}^{-1}(t') \right) \hat{V}_+ \tilde{\rho}^{(0)}(t) \end{align}
Carrying out the integral we find
\begin{align}
I_1 &\approx \hat{V}_- e^{-i \hat{H}_{\rm NH} t} \left[\left(i \hat{H}_{\rm NH}\right)^{-1} e^{i \hat{H}_{\rm NH} t'} \right]^t_0 \hat{V}_+ \tilde{\rho}^{(0)}(t) \nonumber \\ 
&\approx \hat{V}_- \left(i \hat{H}_{\rm NH}\right)^{-1} \hat{V}_+ \tilde{\rho}^{(0)}(t).
\label{groundresult1}
\end{align}
In the last step we have used that the term emerging from the lower limit of the integral at $t'=0$ maintains its time dependence of $e^{- i \hat{H}_{\rm NH} t}$ and is therefore detuned with respect to the term originating from the integral limit at $t'=t$, i.e., by an approximation similar to the rotating wave approximation we keep the unity term in the expression $1-\exp(-i \hat{H}_{\rm NH}t)$.
This condition is equivalent to the standard approximation of adiabatic elimination and is justified provided that the time evolution of the ground states is slow compared to the time scale set by $\hat{H}_{\rm NH}^{-1}$.
The second term of Eq. (\ref{pluggedin}) is treated accordingly, yielding the Hermitian conjugate of the result in Eq. (\ref{groundresult1}).

For the last two Lindblad-type terms in Eq. (\ref{pluggedin}) we carry out the double integral $I_2$. To this end we approximate $\tilde{\rho}^{(0)} (t'')$ in Eq. (\ref{integral2}) by $\tilde{\rho}^{(0)} (t)$. This can be argued the following way: Above we have assumed that the density matrix of the ground states $\tilde{\rho}^{(0)} (t)$ evolves slowly and to second-order in $\hat{V}$. Another dependence on $\tilde{V}(t')$ and $\tilde{V}(t'')$ would only involve features of fourth order in the evolution. We neglect these higher orders by dropping the dependence of $\tilde{\rho}^{(0)} (t')$ and $\tilde{\rho}^{(0)} (t'')$ on the time scales of $\tilde{V}(t')$ and $\tilde{V}(t'')$, which yields $\tilde{\rho}^{(0)} (t)$. Thus, we can separate the integral and write
\begin{align}
I_2 &\approx \frac{1}{2} \left(\int_0^t dt' \tilde{O}^{-1}(t') \right) \hat{V}_+ \tilde{\rho}^{(0)} (t) \hat{V}_- \int_0^{t} dt' (\tilde{O}^{-1})^\dagger(t') \nonumber \\
\label{excitedresult}
&\approx \frac{1}{2} \left(i \hat{H}_{\rm NH}\right)^{-1} \hat{V}_+ \tilde{\rho}^{(0)}(t) \hat{V}_- \left(- i \hat{H}^\dagger_{\rm NH}\right)^{-1}.
\end{align}
Again we have assumed that the ground states are slowly varying compared to the time scale of $\hat{H}_{\rm NH}^{-1}$ so that $\hat{O}(t) P_{\rm g} \simeq P_{\rm g}$ and discarded detuned terms. The remaining term of Eq. (\ref{pluggedin}) yields the same result as Eq. (\ref{excitedresult}).

We insert Eqs. (\ref{groundresult1}) and (\ref{excitedresult}) back into Eq. (\ref{pluggedin}) and transform back into the Schr\"odinger picture. In doing so, we obtain the effective unitary and dissipative dynamics of the ground states,
\begin{align}
P_{\rm g} \dot{\rho}^{(2)} P_{\rm g} = &- i \left(\hat{H}_{\rm eff} - \frac{i}{2} \sum_k (\hat{L}^k_{\rm eff})^{\dagger} \hat{L}_{\rm eff}^k \right) \rho^{(0)} \nonumber + H. c. + \\ &+ \sum_k \hat{L}_{\rm eff}^{k} \rho^{(0)} (\hat{L}_{\rm eff}^{k})^{\dagger}.
\label{effectivegroundresult}
\end{align}
with an effective Hamiltonian and effective Lindblad operators as defined in Eqs. (\ref{effectivehamilton2}) and (\ref{effectivelindblad2}). To reach this form we have used the equality
\begin{align}
\sum_k (\hat{L}^k_{\rm eff})^{\dagger} \hat{L}_{\rm eff}^k &= \nonumber \hat{V}_- (\hat{H}_{\rm NH}^{-1})^{\dagger} \left(\sum_k \hat{L}_k^{\dagger} \hat{L}_k \right) \hat{H}_{\rm NH}^{-1} \hat{V}_+ \nonumber \\
&= - i \hat{V}_- \left(\hat{H}_{\rm NH}^{-1} - (\hat{H}_{\rm NH}^{-1})^{\dagger}\right) \hat{V}_+. \nonumber
\end{align}
From here it can easily be seen that Eq. (\ref{effectivegroundresult}) is equivalent to Eq. (\ref{effectivemaster}). 
Thus, we have reduced the unitary and dissipative dynamics of the open quantum system described by Eq. (\ref{master}) to the effective master equation of Lindblad form in Eq. (\ref{effectivemaster}), obtaining the effective Hamiltonian $\hat{H}_{\rm eff}$ and Lindblad operators $\hat{L}^k_{\rm eff}$ of Eqs. (\ref{effectivehamilton2}) and (\ref{effectivelindblad2}).

The effective Hamiltonian of Eq. (\ref{effectivehamilton2}) is the same as the original result of Feshbach \cite{Feshbach}. In addition, we have found effective Lindblad operators for second-order decay processes. As can be seen from Eq. (\ref{effectivelindblad2}), each of them consists of weak coherent excitation $\hat{V}_+$, evolution between the excited states by a ``propagator" $\hat{H}_{\rm NH}^{-1}$, and subsequent decay $\hat{L}_k$. Thus, adiabatic elimination of the excited states of an open quantum system is formalized in a compact manner by Eqs. (\ref{effectivemaster}--\ref{hnh}).

\begin{figure}[t]
\centering
\includegraphics[width=2.5cm]{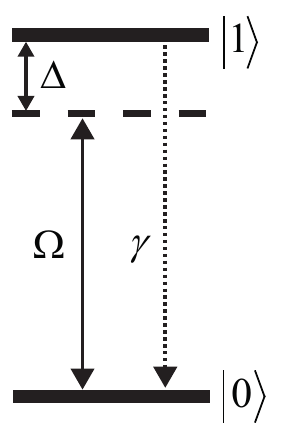}
\caption{Two-level system. A ground state $\ket{0}$ is coherently coupled to an excited state $\ket{1}$ by a field with a Rabi frequency of $\Omega$. The excited state $\ket{1}$ is subject to spontaneous decay at a rate of $\gamma$.}
\label{twolevel}
\end{figure}

\subsection{Example: The two-level system}
\label{Two}
The most elementary configuration our formalism can be applied to is given by a two-level system with a ground state $\ket{0}$ coherently coupled to a spontaneously decaying excited state $\ket{1}$ (Fig. \ref{twolevel}). This example is rather straightforward also without the theory developed here; yet, we include it to demonstrate the general formalism. The Hamiltonian for this system reads
\begin{align}
&\hat{H}=\hat{H}_{\rm e} + \hat{V}\\
&\hat{H}_{\rm e}=\Delta\ket{1}\bra{1}, \hat{H}_{\rm g}=0\\
&\hat{V}=\frac{\Omega}{2}\left(\ket{0}\bra{1}+\ket{1}\bra{0}\right).
\end{align}
Here, $\ket{0}$ and $\ket{1}$ are coupled with a Rabi frequency of $\Omega$ and a detuning of $\Delta$. We write the (de-) excitation as $\hat{V}_+=\frac{\Omega}{2}\ket{1}\bra{0}$ ($\hat{V}_-=\frac{\Omega}{2}\ket{0}\bra{1}$). Spontaneous emission from the excited to the ground level at a rate of $\gamma$ is represented by the Lindblad operator,
\begin{equation}
\hat{L}=\sqrt{\gamma}\ket{0}\bra{1}.
\end{equation}
Consequently, the non-Hermitian Hamiltonian is found to be
\begin{align}
\hat{H}_{\rm NH}&= \tilde{\Delta} \ket{1}\bra{1},
\end{align}
with a complex energy $\tilde{\Delta} \equiv \Delta-\frac{i\gamma}{2}$ that combines the detuning $\Delta$ and the decay width $\gamma$ of the excited state $\ket{1}$ to a single complex quantity $\tilde{\Delta}$.

By applying the effective Hamiltonian formula of Eq. (\ref{effectivehamilton2}) we adiabatically eliminate the excited state $\ket{1}$ and obtain an effective Hamiltonian for the ground state $\ket{0}$:
\begin{align}
\label{effHtwo}
\hat{H}_{\rm eff} &=-\frac{1}{2}\left(\frac{\Omega}{2}\ket{0}\bra{1}\right)\left(\frac{\ket{1}\bra{1}}{\Delta-\frac{i \gamma}{2}} + H. c. \right) \left(\frac{\Omega}{2}\ket{1}\bra{0}\right) \nonumber\\
&=-\frac{\Omega^2 \Delta}{4 \Delta^2+\gamma^2} \ket{0}\bra{0} \equiv \Delta_{\rm eff} \ket{0}\bra{0}.
\end{align}
This effective Hamiltonian describes an effective ac Stark shift $\Delta_{\rm eff}$ of level $\ket{0}$ caused by the coherent driving. By applying Eq. (\ref{effectivelindblad2}) together with $\hat{V}$, $\hat{H}_{\rm NH}$, and $\hat{L}_\gamma$ as specified above, we obtain a single effective Lindblad operator
\begin{align}
\hat{L}_{\rm eff}^\gamma&=\left(\sqrt{\gamma}\ket{0}\bra{1}\right)\left(\frac{1}{\tilde{\Delta}}\ket{1}\bra{1}\right)\left(\frac{\Omega}{2}\ket{1}\bra{0}\right) \nonumber \\
&=\frac{\sqrt{\gamma} \Omega}{2 \Delta-i \gamma} \ket{0}\bra{0}.
\end{align}
The effective scattering rate is thus given by
\begin{align}
\gamma_{\rm eff} \equiv |\bra{0} \hat{L}_{\rm eff} \ket{0}|^2=\frac{\gamma \Omega^2}{4\Delta^2+\gamma^2}.
\end{align}
The effective Lindblad operator $\hat{L}_{\rm eff}$ describes Rayleigh scattering, i.e., elastic scattering of incident laser photons by the transition $\ket{0} \leftrightarrow \ket{1}$. Seen from the atom this effect will contribute to the effective dynamics of state $\ket{0}$ not as a decay but as a dephasing of potential coherent couplings to other states.
\\

Above we have given a rather simple example which could also be easily solved without these techniques. For more complicated situations the formalism developed here is highly useful. In particular, in the following we consider a more complicated scheme involving four levels which is relevant for dissipative state preparation by engineered decay.

\begin{figure}[t]
\centering
\includegraphics[width=6cm]{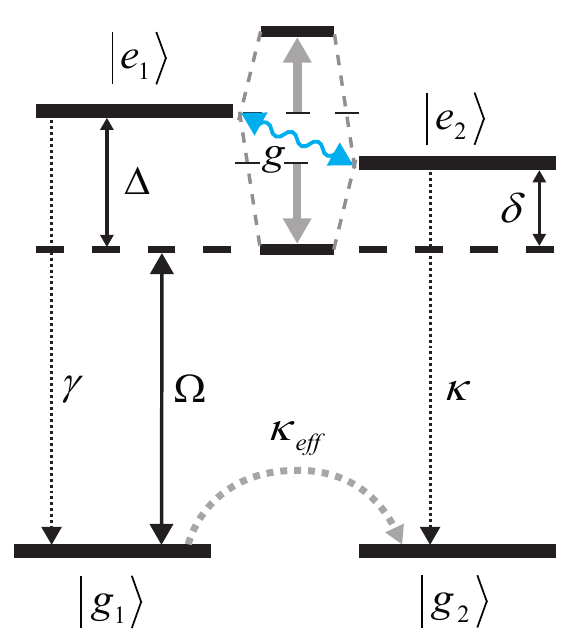}
\caption{(Color online) Dissipative state preparation in a four-level system. An effective decay process consisting of weak excitation from ground state $\ket{g_1}$ with a resonant Rabi frequency of $\Omega$, evolution between the excited levels $\ket{e_1}$ and $\ket{e_2}$, coupled with $g$, and subsequent decay $\kappa$ prepares ground state $\ket{g_2}$. The effective decay rate $\kappa_{\rm eff}$ is engineered by the choice of the detunings of the excited states, $\Delta$ and $\delta$; this depends on how close the dressed states of $\ket{e_1}$ and $\ket{e_2}$ are shifted into resonance with the driving ($\Omega$) by their coupling ($g$).}
\label{FigEngineered}
\end{figure}

\subsection{Engineered decay}
\label{SectionEngineered}
Our formalism allows us to engineer schemes that serve the purpose of dissipative state preparation, such as Refs. \cite{RKS, KRS}. Here, the goal is to prepare a certain desired steady state as the outcome of the evolution of the open system. This is done by tailoring its dissipative dynamics. To this end we use the effective operators of Eqs. (\ref{effectivehamilton2}) and (\ref{effectivelindblad2}) to take nontrivial interactions between the excited states into account and to identify effective decay processes of the system. By an appropriate choice of the system parameters these processes can be engineered by tailoring the ``propagator" $\hat{H}_{\rm NH}^{-1}$.

An example of this is depicted in Fig. \ref{FigEngineered}, showing a four-level system consisting of two ground states $\ket{g_1}$, $\ket{g_2}$ and two excited states $\ket{e_1}$, $\ket{e_2}$. The excited-state Hamiltonian
\begin{align}
\hat{H}_{\rm e}=\Delta\ket{e_1}\bra{e_1}+\delta\ket{e_2}\bra{e_2}+g \left(\ket{e_1}\bra{e_2}+\ket{e_2}\bra{e_1}\right)
\end{align}
contains detunings of $\Delta$ for $\ket{e_1}$ and $\delta$ for $\ket{e_2}$, respectively, and a coupling with a strength of $g$ between $\ket{e_1}$ and $\ket{e_2}$. We assume that there are no processes between the ground states so that $\hat{H}_{\rm g}=0$. The weak classical driving described by
\begin{align}
\hat{V}=\frac{\Omega}{2}\left(\ket{g_1}\bra{e_1}+\ket{e_1}\bra{g_1}\right)
\end{align}
drives the system between $\ket{g_1}$ and $\ket{e_1}$ with a Rabi frequency of $\Omega$. We assume $\ket{e_1}$ to decay to $\ket{g_1}$ at a rate of $\gamma$ and $\ket{e_2}$ to $\ket{g_2}$ at a rate of $\kappa$, represented by the Lindblad operators
\begin{align}
\hat{L}_{\gamma}&=\sqrt{\gamma}\ket{g_1}\bra{e_1},\\
\hat{L}_{\kappa}&=\sqrt{\kappa}\ket{g_2}\bra{e_2}.
\end{align}
The non-Hermitian Hamiltonian is then given by
\begin{align}
\hat{H}_{\rm NH}=&\tilde{\Delta}\ket{e_1}\bra{e_1}+\tilde{\delta}\ket{e_2}\bra{e_2}+g \left(\ket{e_1}\bra{e_2} + H.c. \right),
\end{align}
with complex detunings $\tilde{\delta} \equiv \delta-\frac{i\kappa}{2}$ and $\tilde{\Delta} \equiv \Delta-\frac{i\gamma}{2}$. Using Eqs. (\ref{effectivehamilton2}) and (\ref{effectivelindblad2}) we obtain the effective Hamiltonian and Lindblad operators:
\begin{align}
\hat{H}_{\rm eff}&=-\frac{\Omega^2}{4} {\rm Re}\left(\frac{\tilde{\delta}}{\tilde{\delta} \tilde{\Delta}-g^2}\right) \ket{g_1}\bra{g_1}, \\
\hat{L}^\gamma_{\rm eff}&=\frac{\sqrt{\gamma} \tilde{\delta} \Omega }{2 (\tilde{\delta} \tilde{\Delta} - g^2)} \ket{g_1}\bra{g_1}, \\
\hat{L}^\kappa_{\rm eff}&=\frac{\sqrt{\kappa} g \Omega }{2 (g^2 - \tilde{\delta} \tilde{\Delta})} \ket{g_2}\bra{g_1}.
\end{align}
We note that the effective Hamiltonian $\hat{H}_{\rm eff}$ only contains a shift of $\ket{g_1}$. The effective decay process $\hat{L}_{\rm eff}^\kappa$ effectively prepares the ground state $\ket{g_2}$ from $\ket{g_1}$ at a rate of
\begin{align}
\label{EqEffKappa}
\kappa_{\rm eff}\equiv |\bra{g_2}\hat{L}^\kappa_{\rm eff}\ket{g_1}|^2=\frac{\kappa g^2 \Omega^2}{4 |g^2-\tilde{\delta} \tilde{\Delta}|^2}.
\end{align}
The other process $\hat{L}_{\rm eff}^{\gamma}$ is a dephasing of $\ket{g_1}$ with a rate
\begin{align}
\label{EqEffGamma}
\gamma_{\rm eff}\equiv |\bra{g_1}\hat{L}^\gamma_{\rm eff}\ket{g_1}|^2=\frac{\gamma |\tilde{\delta}|^2 \Omega^2}{4 |\tilde{\delta} \tilde{\Delta}-g^2|^2}.
\end{align}
The strength of the effective Lindblad operator concept is obvious: We immediately derive the effective pumping rates and dynamics of the ground states from the initial operators. If one desires to optimize the preparation of $\ket{g_2}$ from $\ket{g_1}$ which happens at a rate of $\kappa_{\rm eff}$, this can be realized by an appropriate choice of the system parameters, $\Delta$, $\delta$, and eventually, $g$.\\
Let us assume that $\gamma$, $\kappa$, and $g$ are fixed, that the coupling $g$ is strong, $g \gg \gamma, \kappa$, and that the detunings $\Delta$ and $\delta$ are adjustable. Then the optimum is reached by adjusting the detunings to $\delta_{\rm opt}=g^2/\Delta$ and $\Delta_{\rm opt}=g\sqrt{\gamma/\kappa}$, which leads to a maximized effective decay rate of
\begin{align}
\kappa_{\rm eff}^{\rm opt} \approx \frac{\Omega ^2}{8 \gamma}.
\end{align}
\begin{figure}[t]
\centering
\includegraphics[width=8.6cm]{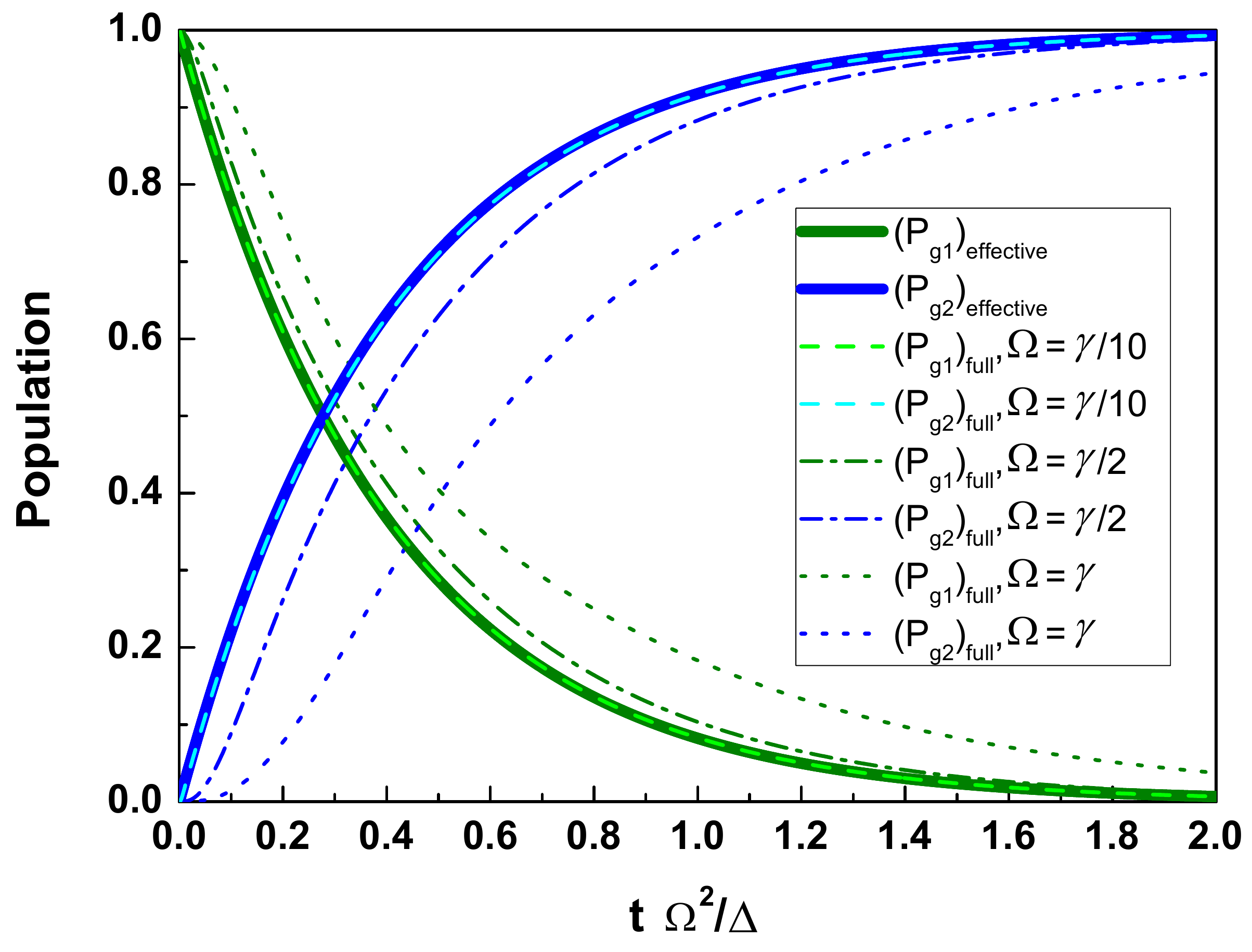}
\caption{(Color online) Effective and full time evolution of a system with engineered decay. The curves illustrate the preparation of the ground state $\ket{g_2}$ (blue lines, starting from $0$) by decay of the ground state $\ket{g_1}$ (green lines, starting from $1$). For weak driving $\Omega \leq \gamma/5$ the effective operators (thick solid) match the full dynamics (dash, $\Omega=\gamma/10$) very accurately. With increasing driving $\Omega$ the assumption of perturbative excitation is no longer valid and the effective evolution deviates from the full dynamics (dash-dot, $\Omega=\gamma/2$; dot, $\Omega=\gamma$). For the simulations the system parameters $\gamma=\kappa=g/10$ and the optimized detunings $\delta=g^2/\Delta$ and $\Delta=g\sqrt{\gamma/\kappa}$ were used.}
\label{FigEngineeredSim}
\end{figure}\\
To compare the effective with the full dynamics, we perform simulations of the evolution of the system by numerically integrating the master equations (\ref{master}) and (\ref{effectivemaster}). The resulting curves are plotted in Fig. \ref{FigEngineeredSim}. Here we show dissipative preparation of $\ket{g_2}$ from $\ket{g_1}$ for the optimal choice of $\Delta$ and $\delta$ as given above. We note that for weak driving $\Omega$ (solid lines, $\Omega=\gamma/10$), the curves of the effective and full dynamics exhibit excellent agreement. For stronger driving $\Omega$ (dash-dot, $\Omega=\gamma/2$; dot, $\Omega=\gamma$), the effective dynamics exhibit increasing deviations. These result from the breakdown of the assumption of weak driving used to derive the effective operators.

\subsubsection{Interpretation and application of the inverse non-Hermitian Hamiltonian $\hat{H}_{\rm NH}^{-1}$}
In general, a good physical understanding of the effective decay mechanisms of an open quantum system is desirable. Even more so, it is essential for developing dissipative state preparation schemes such as the ones in Refs. \cite{KRS, RKS}. Here, engineered decay processes are tailored to prepare a desired steady state. In the following, we discuss the physical meaning of the elements of the inverse non-Hermitian Hamiltonian of the excited-state subspace $\hat{H}_{\rm NH}^{-1}$ for the example at hand. We find that $\hat{H}_{\rm NH}^{-1}$ can be written as
\begin{align}
\label{EqInteracting1}
\hat{H}_{\rm NH}^{-1} = &+\tilde{\Delta}_{\rm eff}^{-1} \ket{e_1}\bra{e_1} + \tilde{\delta}_{\rm eff}^{-1} \ket{e_2}\bra{e_2} \nonumber +\\ &+\tilde{g}_{\rm eff}^{-1} \left(\ket{e_1}\bra{e_2} + \ket{e_2}\bra{e_1} \right),
\end{align}
having defined the quantities
\begin{align}
\label{EqInteracting2}
\tilde{\Delta}_{\rm eff} &\equiv 1/\bra{e_1}\hat{H}_{\rm NH}^{-1}\ket{e_1} = \tilde{\Delta} - g^2/\tilde{\delta} \\
\tilde{\delta}_{\rm eff} &\equiv 1/\bra{e_2}\hat{H}_{\rm NH}^{-1}\ket{e_2} = \tilde{\delta} - g^2/\tilde{\Delta} \\
\tilde{g}_{\rm eff} &\equiv 1/\bra{e_1}\hat{H}_{\rm NH}^{-1}\ket{e_2} = g - \tilde{\delta} \cdot \tilde{\Delta}/g.
\end{align}
Each of the latter quantities can be seen as an effective complex detuning ($\tilde{\Delta}_{\rm eff}$, $\tilde{\delta}_{\rm eff}$) or coupling ($\tilde{g}_{\rm eff}$) of the excited states. $\hat{H}_{\rm NH}^{-1}$ contains their inverses which act as ``propagators" for the effective operators of Eqs. (\ref{effectivehamilton2}) and (\ref{effectivelindblad2}) and therefore govern the strength of the effective processes. Thus, we can also express the effective decay rates of Eqs. (\ref{EqEffKappa}) and (\ref{EqEffGamma}) in terms of the effective complex energies and couplings of the excited states:
\begin{align}
\kappa_{\rm eff} &= \frac{\kappa \Omega^2}{4 |\tilde{g}_{\rm eff}|^2}, \\
\gamma_{\rm eff} &= \frac{\gamma \Omega^2}{4 |\tilde{\Delta}_{\rm eff}|^2}.
\end{align}
We note that our above choice of $\delta=\delta_{\rm opt}$ minimizes $|\tilde{\Delta}_{\rm eff}|^2$, $|\tilde{\delta}_{\rm eff}|^2$, and $|\tilde{g}_{\rm eff}|^2$. Physically, this corresponds to the case where the driving $\hat{V}$ is in resonance with the lower dressed state of the excited states $\ket{e_1}$ and $\ket{e_2}$ (or the upper dressed state for $\delta=-\frac{g^2}{\Delta}$, respectively), as can be seen from Fig. \ref{FigEngineered}. Accordingly, the absolute values of the propagators of $\hat{H}_{\rm NH}^{-1}$ in Eq. (\ref{EqInteracting1}) are maximized under this choice, resulting in an enhanced decay from $\ket{g_1}$ to $\ket{g_2}$. A more detailed account on engineered decay mediated by dressed excited states is provided in Ref. \cite{RKS}.

\section{Extensions of the effective operator formalism}

\subsection{Nonperturbative ground-state coupling}
\label{SectionGround}
The formalism of Eqs. (\ref{effectivemaster}) -- (\ref{effectivelindblad2}) was derived assuming that the ground-state couplings $\hat{H}_{\rm g}$ are much weaker than those of the excited states contained by $\hat{H}_{\rm NH}$. Under this assumption of a perturbative ground-state coupling it was possible to neglect the effect of $\hat{H}_{\rm g}$ on the effective processes.

For strong interactions between the ground states, the action of $\hat{H}_{\rm g}$ can no longer be ignored in the effective processes so that the accuracy of the effective dynamics of Eqs. (\ref{effectivehamilton2}) and (\ref{effectivelindblad2}) will decrease when $\hat{H}_{\rm g}$ approaches $\hat{H}_{\rm NH}$. We will now show how to overcome this drawback by diagonalizing the ground-state Hamiltonian $\hat{H}_{\rm g}$ and including its action in the effective operators.

We can build on the dynamics of the separate subspaces as given by Eqs. (\ref{masterexcited}) and (\ref{pluggedin}), derived without any assumption about the strength of $\hat{H}_{\rm g}$. In contrast to the above derivation, we can no longer assume $\hat{O}(t) P_{\rm g} \simeq \mathds{1} P_{\rm g}$. Since $\hat{H}_{\rm g}$ and $\hat{H}_{\rm NH}$ do not couple the ground and excited subspaces, we can separate the evolution operator into one part for each of the subspaces:
\begin{align}
\label{splitO}
\hat{O}(t) &= P_{\rm g} \hat{O}_{\rm g}(t) P_{\rm g} + P_{\rm e} \hat{O}_{\rm e}(t) P_{\rm e} \nonumber \\
&= e^{-i \hat{H}_{\rm g} t} P_{\rm g} + e^{-i \hat{H}_{\rm NH} t} P_{\rm e}.
\end{align}
We assume that $\hat{H}_{\rm g}$ can be diagonalized
\begin{align}
\label{splitH}
\hat{H}_{\rm g} = \sum_l E_l P_l
\end{align}
with dressed state energies $E_l$ and a projector $P_l \equiv \ket{l}\bra{l}$ for each ground state $l$. Accordingly, we decompose the perturbative excitations $\hat{V}_+$ with respect to the ground states,
\begin{align}
\label{splitV}
\hat{V}_+ = \sum_l \hat{V}_+^{l},
\end{align}
where we have defined $\hat{V}_+^{l} \equiv \hat{V}_+ P_l$ as the excitation from ground state $l$. Given the different energies of the dressed ground states, the effective evolution is now no longer identical for ground states with different energy $E_l$. This additional complication in the integral is taken into account by introducing the sum in Eq. (\ref{splitV}). For instance, for the first term of Eq. (\ref{pluggedin}) we have
\begin{align}
\label{splitthird}
I_1 = P_{\rm g} \tilde{V}(t) \sum_l \int_0^t dt' \ \tilde{V}_l (t) \tilde{\rho}^{(0)}_l (t') P_l
\end{align}
With Eqs. (\ref{splitO})--(\ref{splitV}) this term becomes
\begin{align*}
I_1&= P_{\rm g} \hat{O}_{\rm g}^{-1}(t) \hat{V}_- \hat{O}_{\rm e}(t) \sum_l \int_0^{t} dt' \ e^{i (\hat{H}_{\rm NH}-E_l) t'} \hat{V}_+^l \tilde{\rho}^{(0)}(t) \\ 
&\approx i P_{\rm g} \hat{O}_{\rm g}^{-1}(t) \hat{V}_- \sum_l \left(\hat{H}_{\rm NH}-E_l\right)^{-1} \hat{V}_+^l \hat{O}_{\rm g}(t) \tilde{\rho}^{(0)} (t). 
\end{align*}
The integration is carried out similarly for the other terms in Eq. (\ref{pluggedin}) (the ``sandwich" terms can be neglected for the same reasons as above). Transforming back into the Schr\"odinger picture and arranging the terms as in Eqs. (\ref{effectivehamilton2}) and (\ref{effectivelindblad2}) we find for the effective operators, including nonperturbative ground-state evolution,
\begin{align}
\label{effHground}
\hat{H}_{\rm eff} &= - \frac{1}{2} \left[\hat{V}_- \sum_l \left(\hat{H}_{\rm NH}^{(l)}\right)^{-1} \hat{V}_{+}^{l} + H.c. \right] + \hat{H}_{\rm g}, \\
\label{effLground}
\hat{L}_{\rm eff}^k &= \hat{L}_k \sum_l \left(\hat{H}_{\rm NH}^{(l)}\right)^{-1} \hat{V}_{+}^{l}.
\end{align}
We see that in order to apply this formalism rather than Eqs. (\ref{effectivehamilton2}-\ref{effectivelindblad2}) we replace the general inverse non-Hermitian Hamiltonian $\hat{H}_{\rm NH}$ by initial-state dependent propagators
\begin{align}
\left(\hat{H}_{\rm NH}^{(l)}\right)^{-1} \equiv \left(\hat{H}_{\rm NH}-E_l\right)^{-1}
\end{align}
for each ground state $l$. In doing so we obtain the accurate effective dynamics in the presence of nonperturbative ground-state coupling. In Ref. \cite{RKS} we have used this technique to investigate rapid preparation of entanglement by engineered decay. Below we consider a simpler example of a three-level Raman system.

\begin{figure}[t]
\centering
\includegraphics[width=6cm]{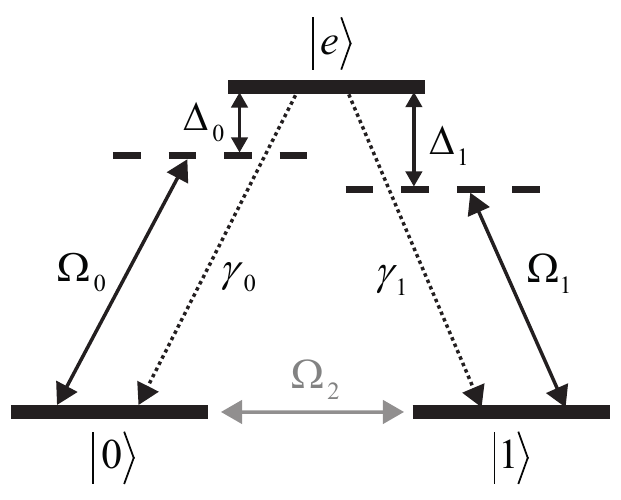}
\caption{The three-level Raman system. Two ground states $\ket{0}$ and $\ket{1}$ are driven up to an excited state $\ket{e}$ with different detunings $\Delta_0$, $\Delta_1$ and Rabi frequencies $\Omega_0$, $\Omega_1$. $\ket{e}$ decays to the ground states via spontaneous emission at rates of $\gamma_0$ and $\gamma_1$. Effects originating from nonperturbative interactions between the ground states (indicated by $\Omega_2$) can also be taken into account by our extended formalism.}
\label{FigureRamanSetup}
\end{figure}

\subsection{Example: The three-level Raman system}
\label{SectionRaman}
A three-level system in Raman configuration is a widely used quantum system so that the understanding of its effective processes is highly relevant. In particular, Ref. \cite{James2} deals with its effective dynamics in the absence of decoherence in great detail. In the following we give a description of the effective dynamics of a three-level Raman system that includes dissipation.

As illustrated in Fig. \ref{FigureRamanSetup}, the system consists of two ground states $\ket{0}$ and $\ket{1}$ and an excited state $\ket{e}$. Coherent driving of the transitions $\ket{0} \leftrightarrow \ket{e}$ and $\ket{1} \leftrightarrow \ket{e}$ is facilitated by two fields, generally with different detunings, $\Delta_0$ and $\Delta_1$, and strengths, $\Omega_0$ and $\Omega_1$. In a time-independent frame the system is described by
\begin{align}
\label{HRaman}
\hat{H}_{\rm g} &= -\Delta_0\ket{0}\bra{0}-\Delta_1\ket{1}\bra{1}, \ \ \hat{H}_{e}=0 \\
\label{VRaman}
\hat{V} &= \frac{\Omega_0}{2}\left(\ket{0}\bra{e}+\ket{e}\bra{0}\right)+\frac{\Omega_1}{2}\left(\ket{1}\bra{e}+\ket{e}\bra{1}\right),
\end{align}
assuming an arbitrarily strong nonperturbative $\hat{H}_{\rm g}$. The time-independent formulation of Eq. (\ref{HRaman}) with detunings $\Delta_0$ and $\Delta_1$ assigned to the ground states allows us to use the formalism of Sec. \ref{SectionGround}. Alternatively, we could load the time dependence on the fields and use the formalism to be presented in Sec. \ref{SectionFields} below. Decay from the excited level $\ket{e}$ into ground states $\ket{0}$ and $\ket{1}$ at rates of $\gamma_0$ and $\gamma_1$ is described by the Lindblad operators
\begin{align}
\hat{L}_{\gamma,0}=\sqrt{\gamma_0}\ket{0}\bra{e} \\
\hat{L}_{\gamma,1}=\sqrt{\gamma_1}\ket{1}\bra{e}.
\end{align}
The non-Hermitian Hamiltonian can be divided into two parts, denoted by the initial state of the exciting field
\begin{align}
\label{hnhraman}
\hat{H}_{\rm NH}^{(0)} = \left(\Delta_0-\frac{i\gamma}{2}\right) \ket{e}\bra{e} \equiv \tilde{\Delta}_0 \ket{e}\bra{e}\\
\label{hnhraman2}
\hat{H}_{\rm NH}^{(1)} = \left(\Delta_1-\frac{i\gamma}{2}\right) \ket{e}\bra{e} \equiv \tilde{\Delta}_1 \ket{e}\bra{e}.
\end{align}
In the last step we have once again assigned complex energies that combine the real detuning and the imaginary decay of the levels.

Using the non-Hermitian Hamiltonians of Eqs. (\ref{hnhraman}) and (\ref{hnhraman2}) together with Eqs. (\ref{effHground})-(-\ref{VRaman}) we obtain the effective Hamiltonian
\begin{align}
\hat{H}_{\rm eff} &= \left(\Delta_0 - \frac{\Delta_0 \Omega_0^2}{4 \Delta_0^2 + \gamma ^2}\right) \ket{0}\bra{0} + \\ &+ \left(\Delta_1 - \frac{\Delta_1 \Omega_1^2}{4 \Delta_1^2 + \gamma ^2}\right) \ket{1}\bra{1} + \nonumber \\ &+
\left(\frac{(\Delta_0+\Delta_1) \Omega_0 \Omega_1}{8 (\Delta_0 - i \gamma/2) (\Delta_1 + i \gamma/2)} \ket{0}\bra{1} + H. c.\right) \nonumber,
\end{align}
where we have defined an overall decay rate of $\gamma=\gamma_0+\gamma_1$. We note that despite the complex terms the effective Hamiltonian is Hermitian. Besides two shift terms similar to the one in Eq. (\ref{effHtwo}), it contains an effective two-photon transition between the two ground states with an effective Rabi frequency of $\tilde{\Omega}_{\rm eff} \equiv \frac{(\Delta_0+\Delta_1) \Omega_0 \Omega_1}{8 (\Delta_0 - i \gamma/2) (\Delta_1 + i \gamma/2)}$.

In the absence of dissipative processes ($\gamma=0$), this effective Hamiltonian equals the time-averaged ground-state Hamiltonian of Gamel and James \cite{James2}, viz.,
\begin{align}
\hat{H}_{\rm eff} = &- \left( \frac{\Omega_0 \Omega_1}{4 \Delta_0} +\frac{\Omega_0 \Omega_1}{4 \Delta_1} \right) \left( \ket{0}\bra{1} + \ket{1}\bra{0} \right) - \nonumber \\ &- \frac{\Omega_0^2}{4 \Delta_0} \ket{0}\bra{0} - \frac{\Omega_1^2}{4 \Delta_1} \ket{1}\bra{1} + \hat{H}_{\rm g}.
\end{align}
Furthermore, we derive the effective Lindblad operators
\begin{align}
\hat{L}^{\gamma,0}_{\rm eff}&=\frac{\sqrt{\gamma_0} \Omega_0}{2 (\Delta_0 - i \gamma/2)} \ket{0}\bra{0}+\frac{\sqrt{\gamma_0} \Omega_1}{2 (\Delta_1 - i \gamma/2)} \ket{0}\bra{1}, \\
\hat{L}^{\gamma,1}_{\rm eff}&=\frac{\sqrt{\gamma_1} \Omega_0}{2 (\Delta_0 - i \gamma/2)} \ket{1}\bra{0}+\frac{\sqrt{\gamma_1} \Omega_1}{2 (\Delta_1 - i \gamma/2)} \ket{1}\bra{1}.
\end{align}
Besides one loop-term for each of the ground states in this setup, these operators contain effective decays from either ground state to the other, the strength of which is given by
\begin{align}
\gamma_{\rm eff}^{0 \rightarrow 1} &\equiv |\bra{1}\hat{L}^{\gamma,1}_{\rm eff}\ket{0}|^2 = \frac{\gamma_1 \Omega_0^2}{4 \Delta_0^2 + \gamma^2}, \\
\gamma_{\rm eff}^{1 \rightarrow 0} &\equiv |\bra{0}\hat{L}^{\gamma,0}_{\rm eff}\ket{1}|^2 = \frac{\gamma_0 \Omega_1^2}{4 \Delta_1^2 + \gamma^2}.
\end{align}
We note that depending on the relative strength of the effective quantities $\Omega_{\rm eff}$ and $\gamma_{\rm eff}$, the resulting effective dynamics will either be governed by coherent or decoherent behavior.
\begin{figure}[t]
\centering
\includegraphics[width=8.6cm]{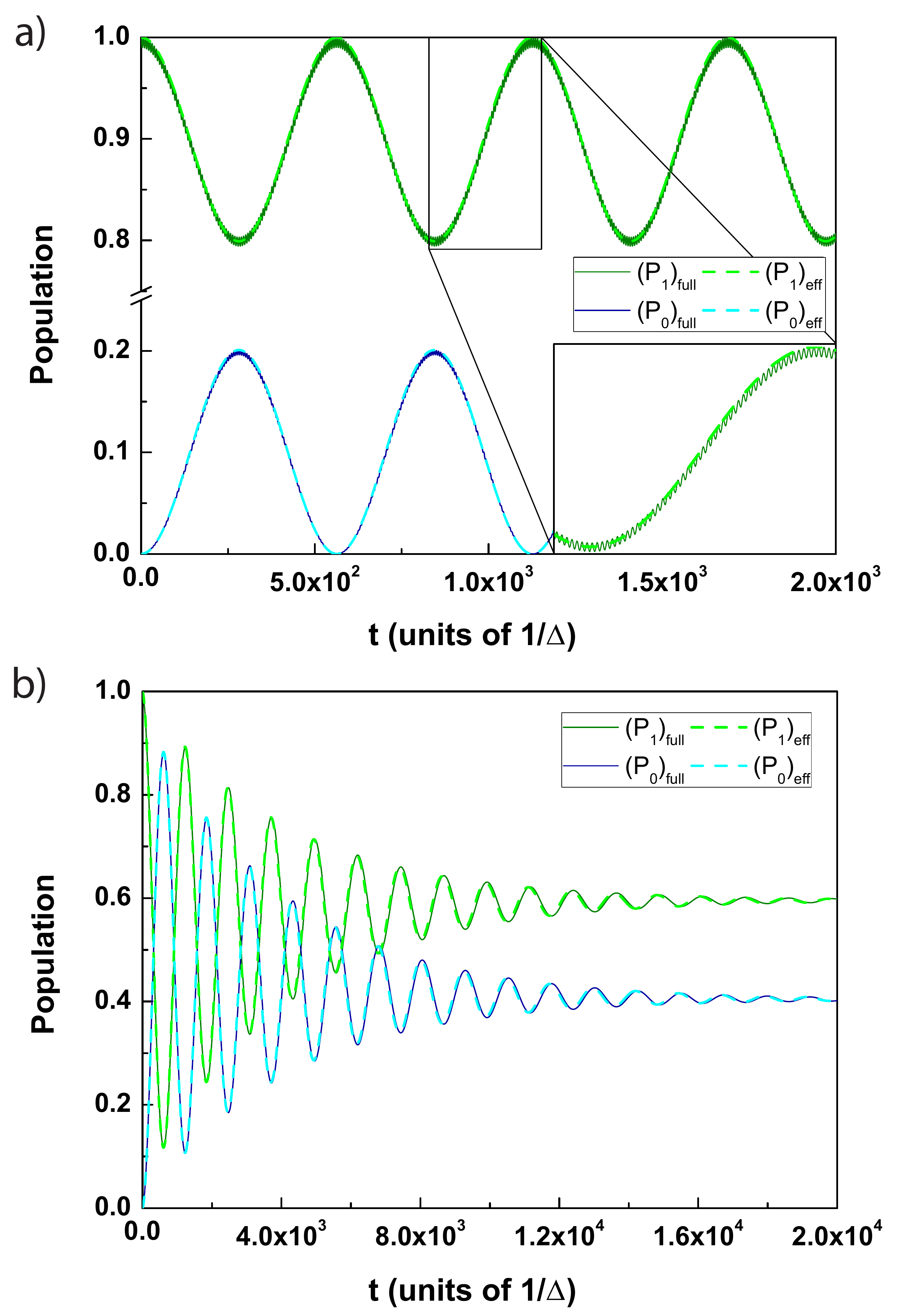}
\caption{(Color online) Comparison of effective and full evolution of a three-level Raman system. Curves obtained from numerical integration of the effective master equation (dashed) agree with results from the full master equation (solid) both (a) in the absence ($\gamma=0$) and (b) in the presence of dissipation ($\gamma \neq 0$). The effective operators are found to model the slow dynamics of the two ground states $\ket{0}$ (blue, starting from $0$) and $\ket{1}$ (green, starting from $1$) very accurately, averaging out the fast oscillations (inset in a). For the simulations we used the parameters $\Omega_{0,1}=\Delta/10$ ($\Delta=\Delta_0+\Delta_1$): (a) $\Delta_1-\Delta_0=\Delta/1000$, $\gamma_{0,1}=0$, (b) $\Delta_1-\Delta_0=\Delta/100$, $\gamma_{0,1}=\Delta/10$.}
\label{FigureRamanSimulation}
\end{figure}
\\
This is visualized in Fig. \ref{FigureRamanSimulation} where we have plotted simulated curves of the full and effective dynamics obtained by numerical integration of the master equations (\ref{master})--(\ref{effectivemaster}). In Fig. \ref{FigureRamanSimulation}(a) we show a purely unitary case, and in \ref{FigureRamanSimulation}(b) a mixed case with both coherent and dissipative processes present.

In the purely unitary case ($\gamma=0$) shown in Fig. \ref{FigureRamanSimulation}(a) we see that the populations of the two ground levels exhibit Rabi oscillations at a high and a low frequency. The high-frequency oscillations correspond to the virtual excitation of the excited state $\ket{e}$. These oscillations are explicitly excluded from the formalism developed here and are therefore not present in the evolution with the effective operators, as can be seen from the inset in Fig. \ref{FigureRamanSimulation}(a). Nevertheless, the formalism captures the slow dynamics of the ground states.

A case including dissipation ($\gamma \neq 0$) is shown in Fig. \ref{FigureRamanSimulation}(b). Here we see that even for Rabi oscillations sweeping almost the entire population between the ground states, the effective dynamics match the time evolution of the full master equation with very high precision. For large times $t$ the oscillations are damped out and the dynamics converge into a steady state.

Finally, we comment on the situation where the ground states are coupled by another field of strength $\Omega_2$, as illustrated in Fig. \ref{FigureRamanSetup}. In case this additional interaction is perturbative it can simply be included in $\hat{H}_{\rm g}$, and hence, in $\hat{H}_{\rm eff}$, without affecting the other terms. In the nonperturbative case the effect of $\Omega_2$ on the effective processes is caught by changing into a frame in which the ground-state Hamiltonian $\hat{H}_{\rm g}$ is diagonal. From there the formalism of Sec. \ref{SectionGround} can be applied.

\subsection{Several perturbations or fields}
\label{SectionFields}
In the following we present an extension of our effective operator formalism to several perturbations or fields $\hat{V}_f$, where $f$ denotes the particular field and $\omega_f$ its frequency. Then we can write the perturbations as
\begin{align}
\hat{V}(t) &= \sum_f \hat{V}_+^{f}(t) + H. c. = \sum_f \hat{v}_+^{f} e^{-i \omega_f t} + H. c.
\end{align}
The formalism we develop can also be used to include nonperturbative ground-state coupling as in Sec. \ref{SectionGround}. Still, at the first glance the assumption of several fields seems problematic: so far we assumed a rotating frame of reference in which $\hat{V}$ is time independent. However, our formalism can be derived without this claim, starting from Eqs. (\ref{masterexcited}) and (\ref{pluggedin}). As opposed to the previous case, where we chose to work in the interaction picture, we now keep the time dependence in the perturbations. For simplicity we choose a time-independent frame with respect to the interactions inside the subspaces ($\frac{\partial \hat{H}_{\rm e}}{\partial t}=\frac{\partial \hat{H}_{\rm g}}{\partial t}=0$). Again, for perturbative ground-state coupling $\hat{H}_{\rm g}$ the ground-state evolution becomes negligible, $\hat{O}_{\rm g}(t) \approx \mathds{1}$. The perturbative evolution of the ground states turns into a sum for the different fields $f$:
\begin{align}
\label{drivethird}
I_1 &= P_{\rm g} \tilde{V}(t) \int_0^t dt' \ \tilde{V}(t') \tilde{\rho}^{(0)}(t') P_{\rm g} = \\
&\approx P_{\rm g} \hat{V}_-(t) \hat{O}_{\rm e}(t) \sum_f \int_0^{t} dt' \ e^{i (\hat{H}_{\rm NH}-\omega_f) t'} \hat{v}_+^f P_k \rho^{(0)} (t) \nonumber \\
&\approx i P_{\rm g} \hat{V}_-(t) \sum_f \left(\hat{H}_{\rm NH}-\omega_f\right)^{-1} \hat{V}_+^f (t) P_{\rm g} \rho^{(0)} (t) \nonumber.
\end{align}
We then obtain the effective operators
\begin{align}
\label{effHfield}
\hat{H}_{\rm eff} &= - \frac{1}{2} \left[\hat{V}_-(t) \sum_f \left(\hat{H}_{\rm NH}^{(f)}\right)^{-1} \hat{V}_{+}^{f}(t) + H.c. \right] + \hat{H}_{\rm g}\\
\label{effLfield}
\hat{L}_{\rm eff}^k &= \hat{L}_k \sum_f \left(\hat{H}_{\rm NH}^{(f)}\right)^{-1} \hat{V}_{+}^{f}(t)
\end{align}
with one propagator $(\hat{H}_{\rm NH}^{(f)})^{-1} \equiv (\hat{H}_{\rm NH}-\omega_f)^{-1}$ for each field $f$.

\subsection{General formalism}
\label{SectionGeneral}
In Sec. \ref{SectionGround} we derived an extension of the effective operators in Eqs. (\ref{effectivehamilton2}) and (\ref{effectivelindblad2}) for nonperturbative ground-state coupling and in Sec. \ref{SectionFields} we provided the extension for several fields. In a situation with both aspects present these two extensions can be directly combined. We find the effective operators
\begin{align}
\label{effHall}
\hat{H}_{\rm eff} &= - \frac{1}{2} \left[\hat{V}_- \sum_{f, l} \left(\hat{H}_{\rm NH}^{(f, l)}\right)^{-1} \hat{V}_{+}^{(f, l)}(t) + H.c. \right] + \hat{H}_{\rm g}\\
\label{effLall}
\hat{L}_{\rm eff}^k &= \hat{L}_k \sum_{f, l} \left(\hat{H}_{\rm NH}^{(f, l)}\right)^{-1} \hat{V}_{+}^{(f, l)}(t)
\end{align}
with field- and state-dependent propagators
\begin{align}
\left(\hat{H}_{\rm NH}^{(f, l)}\right)^{-1} \equiv \left(\hat{H}_{\rm NH} - E_l - \omega_f\right).
\end{align}
This quantity contains both the information about the initial state and the exciting field; thus, it is the most general propagator expression presented in this work.

\section{Comparison to other methods}
\label{SectionComparison}
In the following we compare the results obtained here with other methods from the literature. The formalism we have presented here is equivalent to the standard approach of adiabatic elimination in quantum optics (see, e.g., Ref. \cite{Brion}) and is essentially a formalization of it. In most approaches adiabatic elimination is done at the level of equations of motion. This procedure can therefore be rather tedious, as it requires the derivation of the equations of motion followed by various manipulations of the equations, which are then often used to extract effective operators. For comparison our formalism works directly at the operators and immediately gives the effective operators without reverting to the equations of motion.

The effective operators of Eqs. (\ref{effectivehamilton2}) and (\ref{effectivelindblad2}) are the most compact formalism presented in this work. For the assumption of perturbative ground-state coupling and a single perturbative exciting field these operators match the time evolution very precisely. The extended operators of Eqs. (\ref{effHground}) and (\ref{effLground}), (\ref{effHfield}) and (\ref{effLfield}), and (\ref{effHall}) and (\ref{effLall}) allow for the same precision, but also in the presence of nonperturbative ground-state interactions and several exciting fields.

As could be seen from the example of the three-level Raman scheme in Sec. \ref{SectionRaman}, in the absence of decoherence the effective Hamiltonian method of James and co-workers \cite{James1, James2} can lead to similar results as our formalism. However, as opposed to Ref. \cite{James2}, we do not find any additional decoherence terms emerging from averaging over the fast coherent evolution of the fields.

Also, the Feshbach projection-operator method of Refs. \cite{Feshbach, Rotter} allows for descriptions of open system dynamics by means of an effective non-Hermitian Hamiltonian. We see that if we ignore the feeding term $\sum_k \hat{L}^k_{\rm eff} \rho^{(0)} (\hat{L}_{\rm eff})^\dagger$ in Eq. (\ref{effectivegroundresult}) the evolution is described by an effective non-Hermitian Hamiltonian 
\begin{align}
\hat{H}_{\rm eff, NH} &= \hat{H}_{\rm eff} - \frac{i}{2} \sum_k (\hat{L}^k_{\rm eff})^{\dagger} \hat{L}_{\rm eff}^k \\ &= \hat{V}_- \hat{H}_{\rm NH}^{-1} \hat{V}_+ + \hat{H}_{\rm g}.
\end{align}
This Hamiltonian is equivalent to the one of Refs. \cite{Feshbach, Rotter}. In the language of the Monte Carlo wave function method \cite{Molmer}, $\hat{H}_{\rm eff, NH}$ accounts for the ``no-jump" evolution of the ground states. In contrast to this method, our effective formalism goes beyond including effects of non-Hermitian time evolution in the effective Hamiltonian, as we also include the feeding term. As a result we separate the non-Hermitian effective evolution into a Hermitian part with a (Hermitian) Hamiltonian $\hat{H}_{\rm eff}$ and a non-Hermitian part with effective Lindblad operators $\hat{L}^k_{\rm eff}$. 

\section{Conclusion and outlook}
We have presented an effective operator method covering both coherent Hamiltonian and dissipative Lindblad interactions. Our effective operator formalism allows us to reduce the complexity of an open quantum system considerably by restricting its time evolution to an effective master equation describing ground-state to ground-state processes.

Our effective operator formalism is useful for understanding the quantum dynamics of complex open systems by identifying their effective dissipative interactions and by reducing high-dimensional evolution to an effective master equation of the ground states. More specifically, our effective operators can be applied to identify and tailor effective decay processes involving coherent driving and naturally occurring sources of noise in open quantum systems. In particular, we have found the presented operators to allow for the development of physical schemes for dissipative quantum computing and dissipative state engineering \cite{KRS, RKS}.

\section{Acknowledgements}
We thank Michael Kastoryano, David Reeb, Chris Pethick, and Bernhard Mehlig for discussions, and Emil Zeuthen for reading the manuscript. This work was supported by the Villum Kann Rasmussen Foundation and the Danish National Research Foundation. F.R. acknowledges support from the German Academic Exchange Service (DAAD).

\end{document}